\documentstyle[preprint,aps,epsfig,amssymb]{revtex}
\tighten
\begin{document}
\draft
\preprint{Preprint Numbers: \parbox[t]{45mm}{KSU-CNR-104-99\\
                                             nucl-th/9905056}}

\title{Bethe--Salpeter Study of Vector Meson Masses \\ and Decay Constants}

\author{Pieter Maris and Peter C. Tandy} 
\address{Center for Nuclear Research, Department of Physics,\\ 
         Kent State University, Kent OH 44242}
\date{\today}
\maketitle
\begin{abstract}
The masses and decay constants of the light vector mesons
$\rho/\omega$, $\phi$ and $K^\star$ are studied within a ladder-rainbow
truncation of the coupled Dyson--Schwinger and Bethe--Salpeter equations
of QCD with a model 2-point gluon function.  The approach is consistent
with quark and gluon confinement, reproduces the correct one-loop
renormalization group behavior of QCD, generates dynamical chiral
symmetry breaking, and preserves the relevant Ward identities.  The one
phenomenological parameter and two current quark masses are fixed by
requiring that the calculated $f_\pi$, $m_\pi$ and $m_K$ are correct.
The resulting $f_K$ is within 3\% of the experimental value.  For the
vector mesons, all eight transverse covariants are included and the
dominant ones are identified; the complete angle dependence of the
amplitudes is also retained.  The calculated values for the masses
$m_\rho$, $m_\phi$ and $m_{K^\star}$ are within 5\%, while the 
decay constants $f_\rho$, $f_\phi$ and $f_{K^\star}$ for electromagnetic
and leptonic decays are within 10\% of the experimental values.

\end{abstract}

%
\section{Introduction}

A realistic description of vector mesons at the quark-gluon level is an
important element in advancing our understanding of hadron dynamics and
reaction processes at scales where QCD degrees of freedom are relevant.
They are easily produced as decay products in electro-excitation of
baryon resonances and also as precursors to di-lepton events in
relativistic heavy-ion collisions.  Flavorless vector mesons couple
directly to the photon and play an important role in the phenomenology
of electromagnetic coupling to hadrons.  This is exemplified by the
general phenomenological success of the Vector Meson Dominance model
which assumes that the electromagnetic current is saturated by the
vector mesons. The ground state vector mesons, being spin modes, sample
the \mbox{$\bar q q$} bound state dynamics in a way that is
complementary to that of the ground state pseudoscalar (PS) mesons.
They also explore quark and gluon confinement since the vector masses
are greater than the sum of typical constituent quark masses.  Mesonic
strong decays such as \mbox{$\rho \rightarrow \pi\pi$} and \mbox{$\phi
\rightarrow K K$}, radiative decays such as \mbox{$\rho \rightarrow
\pi\gamma$} and \mbox{$K^\star \rightarrow K \gamma$}, and electromagnetic
decays such as \mbox{$\rho^0 \rightarrow e^+e^-$} and \mbox{$\phi
\rightarrow e^+e^-$} can probe aspects of the underlying quark-gluon
dynamics that are complementary to what is learnt from PS mesons.

The PS mesons, especially the pion and kaon, have for a long time been a
major focus of attempts to understand the internal structure of hadrons
from nonperturbative QCD.  Chiral symmetry provides an assistance in the
PS case that is not available to other mesons like the vectors we
discuss in this paper.  In the chiral limit of massless QCD, the
dynamical breaking of chiral symmetry generates masses for the light
quarks that are consistent with the empirical constituent masses deduced
from the hadronic spectrum.  The lowest PS mesons, which would be
massless in this limit as dictated by the Goldstone theorem, acquire a
mass through the explicit breaking of chiral symmetry induced by the
current quark masses.  This phenomenon dominates the systematics of the
ground state PS octet and provides chiral Ward identities that relate
dynamical quantities in a way that simplifies somewhat the task of
modeling low energy QCD.  For example, the axial Ward identity dictates
that the chiral limit Bethe--Salpeter (BS) amplitude for a pseudoscalar
$\bar q q$ bound state in the dominant $\gamma_5$ channel is given by
\mbox{$B_0(p^2)/f_P$} where $B_0$ is the scalar part of the chiral quark
self-energy, and $f_P$ is the meson weak decay constant.  Consideration
of the symmetry breaking effect of current masses within the PS bound
state dynamics leads to an exact formula for $m_P$ the PS meson
mass~\cite{MRT98}.  One corollary is the Gell-Mann--Oakes--Renner relation 
at small current quark masses where \mbox{$m_P \propto \sqrt{m_q}$}; a
second corollary is the behavior \mbox{$m_P \propto m_Q$} for heavy quark
PS mesons~\cite{IKR99}.
With these aids, an efficient and qualitatively useful phenomenology for
observables and other quantities associated with the pion and kaon can
be produced without explicit solution of the bound state Bethe--Salpeter
equation (BSE).  The only dynamical input required is the dressed quark
propagator as defined by the quark Dyson--Schwinger equation (DSE).

The study of hadronic processes is often facilitated by parameterizing
DSE quark propagator solutions into analytic forms with a few parameters
readjusted to accommodate chiral observables~\cite{R96}.  With such an
approach, $m_{\pi/K}$ and $f_{\pi/K}$, the charge radii $r_{\pi/K}$ and
form factors $F_{\pi/K}(Q^2)$, and the $\pi-\pi$ scattering lengths have
been well described~\cite{R96}.  This approach has also been successful
in studies of coupling constants and form factors for processes such as
\mbox{$\pi^0\rightarrow\gamma\gamma$}~\cite{FMRT95}, and
\mbox{$\gamma\pi\rightarrow \pi\pi$}~\cite{AR96}.  The incorporation of
vector mesons has been hindered by the lack of a symmetry-based means of
obtaining approximate dynamical insight without direct solution of the
vector BSE.  There is a conserved current and a corresponding vector
Ward--Takahashi identity linking the longitudinal vector vertex with the
quark propagator.  However, this Ward identity does not constrain the
transverse vector meson BS amplitude and therefore purely
phenomenological vector BS amplitudes have often been used to study
processes involving vector mesons in this framework.  Successful
applications of this type include the decays \mbox{$\rho\rightarrow
\pi\pi$} and \mbox{$\rho\rightarrow\pi\gamma$}~\cite{pctrev,HP98} and
diffractive electroproduction of vector mesons~\cite{PichLee}.
Uncertainties concerning contributions to hadronic observables from the
neglected covariants in the PS meson BS amplitudes beyond the canonical
$\gamma_5$ were not addressed until recently~\cite{sep97,MR97}.  Several
studies~\cite{sep97,FR96} also incorporated some aspects of vector BSE
dynamics and were able to make a crude assessment of the role of
sub-dominant covariants of the rho~\cite{Tdubrov98} in
\mbox{$\rho\rightarrow\pi\pi$} and \mbox{$\rho \rightarrow \pi\gamma$}
but at the expense of using a separable Ansatz~\cite{sep97} for the BSE
kernel.  A recent study of heavy meson decays also employs
phenomenological BS amplitudes for vector mesons~\cite{IKR99}.

The persistent outcome of the above studies is that soft observables
associated with the pion and kaon are consistently and naturally
described in terms of the momentum dependent quark self-energy from
realistic solutions~\cite{dserev} of the quark DSE.  Important to the
success of this approach are the features of quark confinement and
dynamical chiral symmetry breaking that are implemented through a strong
enhancement in the infrared behavior of the effective quark-quark
interaction (or effective gluon 2-point function~\cite{BP89}) in rainbow
approximation.  In the light pseudoscalar sector, the most comprehensive
and quantitatively reliable study to date~\cite{MR97} involves direct
solution of the bound state BSE in conjunction with quark DSE solutions
for propagators.  That work represents the development of an appropriate
phenomenological representation for the infrared structure of the gluon
2-point function in conjunction with a bare quark-gluon vertex so that
the DSE solution for the quark propagator exhibits dynamical chiral
symmetry breaking as well as confinement~\cite{confinement} and, through
the BSE, produces a good description soft pion and kaon observables.
One of the aims of Ref.~\cite{MR97} was an exposition of the detailed
numerical consequences of the constraint provided by the axial vector
Ward--Takahashi identity (AV-WTI) upon PS meson dynamics.  This
constraint is formally assured by the coordinated rainbow-ladder
truncation of the DSE-BSE complex of DSEs.  The inclusion of all
possible covariants for the BS amplitude was found necessary to
numerically preserve the constraint and to obtain quantitatively
accurate observables.  Of general importance are two other advances
represented by Ref.~\cite{MR97}.  Firstly, since the ultraviolet
structure of the model is equipped with the one-loop renormalization
group properties from perturbative QCD, it is a realistic and covariant
hadron model that can be unambiguously evolved in scale.  Secondly, it
is produced by well-defined truncations of the QCD equations of motion
(DSEs), and thus can be systematically improved by including
higher-order corrections to the quark-antiquark scattering kernel.

The extension of the DSE approach to vector mesons is explored here.
Solution of the vector BSE is more difficult than in the PS case because
of the significantly larger number of covariants that must be
investigated and also because the higher masses produce a larger domain
of the quark complex $p^2$ plane that must be sampled.  This latter
issue was avoided in a previous work~\cite{JM93} that made an extensive
study of the meson spectrum from the ladder-rainbow truncation of the
DSE-BSE system.  In that approach, a derivative expansion of the quark
self-energy was used to infer the behavior away from the real axis and
some attempt was made to estimate the resulting error.  The implications
for quark confinement in that approach are unclear.  One of our aims
here is to generate vector meson BS amplitudes without compromising the
analytic structure in a way that may impair the subsequent explorations
of meson decays and form factors.  These amplitudes can then be used to
calibrate and guide approximate representations that simplify the study
of hadronic interactions.

In this paper we calculate the ground state vector mesons $\rho/\omega$,
$K^\star$ and $\phi$ in the DSE-BSE approach, using the ladder-rainbow
truncation.  The effective quark-quark interaction is fixed by pion and
kaon properties and we investigate the quality of generated vector meson
masses and decay constants.  In Sec. II we outline the framework of the
DSE approach we employ along with the truncation and the Ansatz we use
to specify the kernel (or effective gluon 2-point function) for both the
quark DSE and the bound state BSE.  Our investigations are conducted
with a variation of the kernel Ansatz that was developed in
Ref.~\cite{MR97} for the pion and kaon.  To facilitate the analysis and
solution of the vector BSE, we have employed a convenient set of eight
Dirac covariants that satisfy both the CPT constraints and a
trace-orthogonality property.  These are presented and discussed in
Sec. III.  Also in that Section we outline the technique of expansion of
the amplitudes in terms of Chebyshev polynomials that is sometimes used
to resolve the angle dependence and reduce the BSE to a set of
one-dimensional equations.  In Sec. IV the meson decay constants treated
here are defined.  Results are presented and discussed in Section V, and
a summary and conclusion follows in Sec. VI.  Some technical details are
collected into an Appendix.

\section{Dyson--Schwinger equations}

In a Euclidean space formulation, with
$\{\gamma_\mu,\gamma_\nu\}=2\delta_{\mu\nu}$, $\gamma_\mu^\dagger =
\gamma_\mu$ and $a\cdot b=\sum_{i=1}^4 a_i b_i$, the DSE for the
renormalized dressed-quark propagator is
\begin{eqnarray}
\label{gendse}
 S(p)^{-1} & = & Z_2\,i\gamma\cdot p + Z_4\,m(\mu)
        + Z_1 \int^\Lambda_q \,g^2 D_{\mu\nu}(p-q) 
        \frac{\lambda^a}{2}\gamma_\mu S(q)\Gamma^a_\nu(q,p) \,,
\end{eqnarray}
where $D_{\mu\nu}(k)$ is the renormalized dressed-gluon propagator,
$\Gamma^a_\nu(q;p)$ is the renormalized dressed-quark-gluon vertex, and
$\int^\Lambda_q \equiv \int^\Lambda d^4 q/(2\pi)^4$ represents
mnemonically a translationally-invariant regularization of the integral,
with $\Lambda$ the regularization mass-scale.  The final stage of any
calculation is to remove the regularization by taking the limit $\Lambda
\to \infty$.  The solution of Eq.~(\ref{gendse}) has the general form
\begin{equation}
\label{sinvp}
S(p)^{-1} = i \gamma\cdot p A(p^2,\mu^2) + B(p^2,\mu^2)\,,
\end{equation}
and the renormalization condition is
\begin{equation}
\label{renormS}
S(p)^{-1}\bigg|_{p^2=\mu^2} = i\gamma\cdot p + m(\mu)\,,
\end{equation}
at a sufficiently large spacelike $\mu^2$, with $m(\mu)$ the
renormalized quark mass at the scale $\mu$.  The renormalization
constants for the quark-gluon-vertex, the quark wave-function, and the
mass, namely $Z_1(\mu^2,\Lambda^2)$, $Z_2(\mu^2,\Lambda^2)$ and
$Z_4(\mu^2,\Lambda^2)$ respectively, depend on the renormalization point
and the regularization mass-scale.  In Eq.~(\ref{gendse}), $S$,
$\Gamma^a_\mu$ and $m(\mu)$ depend on the quark flavor, although we have
not indicated this explicitly.  However, in our analysis we assume, and
employ, a flavor-independent renormalization scheme and hence all the
renormalization constants are flavor-independent.

\subsection{Meson Bethe--Salpeter equation}

The renormalized, homogeneous BSE for a bound state of a quark of flavor
$a$ and an antiquark of flavor $b$ having total momentum $P$ is given by
\begin{equation}
 \Gamma^{ab}_M(p;P) =  \int^\Lambda\!\frac{d^4q}{(2\pi)^4} K(p,q;P) 
        S^a(q+\eta P) \Gamma^{ab}_M(q;P) S^b(q-\bar\eta P)\;, 
\label{matbse}
\end{equation}
where \mbox{$\eta + \bar\eta = 1$} describes momentum sharing,
$\Gamma^{ab}_M(p;P)$ is the BS amplitude, and $M$ specifies the meson
type: pseudoscalar, vector, axial-vector, or scalar.  In this paper we
consider the pseudoscalar and vector amplitudes only.  The kernel $K$
operates in the direct product space of color and Dirac spin for the
quark and antiquark and is the renormalized, amputated $\bar q q$
scattering kernel that is irreducible with respect to a pair of $\bar q
q$ lines.  It is often convenient to express Eq.~(\ref{matbse}) in the
abbreviated form
\begin{eqnarray}
\label{genbse}
        \left[\Gamma^{ab}_M(p;P)\right]_{tu} &= & \int^\Lambda_q  \,
        K^{rs}_{tu}(p,q;P)\,\left[\chi_M^{ab}(q;P)\right]_{sr} \,,
\end{eqnarray}
where \mbox{$\chi_M^{ab}(q;P) := S^a(q_+) \Gamma_M^{ab}(q;P)
S^b(q_-)$} is the BS wave function, $q_+ := q + \eta\, P$, $q_- :=
q - \bar \eta\, P$, and the labels $r$,\ldots,$u$ represent color- and
Dirac-matrix indices.  This equation defines an eigenvalue problem with
physical solutions at the mass-shell points \mbox{$P^2=-m^2$} with $m$
being the bound state mass.

The canonical normalization condition of the solution of the homogeneous
BSE is
\begin{eqnarray}
 2 P_\mu &=& \frac{\partial}{\partial P_\mu} 
        \Bigg\{\int^\Lambda_q {\rm Tr}_{CD}
        \left[\bar\Gamma_M^{ba}(q;-K)\, S^a(q+\eta P)\, 
        \Gamma_M^{ab}(q;K)\, S^b(q-\bar\eta P)\right]  +  
\nonumber  \\ & & \;\;\;\;\;\;\; \left. 
        \int^\Lambda_{q}\int^\Lambda_{k} 
        \left[\bar\chi_M^{ba}(k;-K)\right]_{ut} \,
        K^{rs}_{tu}(k,q;P) \,\left[\chi_M^{ab}(q;K)\right]_{sr} 
        \Bigg\}\right|_{P^2=K^2=-m^2} \,,
\label{gennorm}
\end{eqnarray}
where $\bar \Gamma_M(k,-P)^{\rm t} = C^{-1} \Gamma_M(-k,-P) C$, in which
$C=\gamma_2 \gamma_4$ is the charge conjugation matrix, and $X^{\rm t}$
denotes the matrix transpose of $X$.  The trace in the first term is
over both color and Dirac indices.  If the quark-antiquark scattering
kernel $K$ is independent of the total momentum $P$, as is the case in
the ladder truncation we consider here, then the second term vanishes.

\subsection{Ladder-rainbow truncation}
\label{secladder}

We use a ladder truncation for the BSE 
\begin{equation}
\label{ourBSEansatz}
        K^{rs}_{tu}(p,q;P) \to
        -{\cal G}((p-q)^2)\, D_{\mu\nu}^{\rm free}(p-q)
        \left(\frac{\lambda^a}{2}\gamma_\mu\right)^{ru} \otimes
        \left(\frac{\lambda^a}{2}\gamma_\nu\right)^{ts} \,,
\end{equation}
which is consistent with a rainbow truncation for the quark DSE
\begin{equation}
\label{ourDSEansatz}
Z_1\, \int^\Lambda_q \,
g^2 D_{\mu\nu}(p-q) \frac{\lambda^a}{2}\gamma_\mu S(q)
\Gamma^a_\nu(q,p)
\to
\int^\Lambda_q \,
{\cal G}((p-q)^2)\, D_{\mu\nu}^{\rm free}(p-q)
 \frac{\lambda^a}{2}\gamma_\mu S(q)
\frac{\lambda^a}{2}\gamma_\nu \,.
\end{equation}
Here $D_{\mu\nu}^{\rm free}(k)$ is the perturbative gluon propagator in
Landau gauge.  The model is completely specified once a form is chosen
for the ``effective coupling'' ${\cal G}(k^2)$.  

The consistency of Eqs.~(\ref{ourBSEansatz}) and (\ref{ourDSEansatz})
lies in the fact that the axial-vector Ward--Takahashi identity is
preserved~\cite{MRT98,MR97}.  This ensures that in the chiral limit the
ground state PS mesons are massless even though the quark mass functions
are strongly enhanced in the infrared.  In the physical case of explicit
chiral symmetry breaking, it also ensures an exact relation between the
PS meson mass and weak decay constant, the current quark masses, and the
residue at the PS meson pole in the PS vertex~\cite{MRT98,MR97}.  The
analysis in Ref.~\cite{BRvS96} shows that the next-order contributions
to the kernel in a quark-gluon skeleton graph expansion, have a
significant amount of cancellation between repulsive and attractive
corrections for pseudoscalar mesons.  Indications are that this is also
the case in the vector channel, which strongly supports the use of
ladder truncations in these cases.

In choosing a form for ${\cal G}(k^2)$ we know that the behavior of the
QCD running coupling $\alpha(k^2)$ in the ultraviolet, i.e. for $k^2>
2$-$3\,$GeV$^2$, is well described by perturbation theory.  In
principle, constraints on the infrared form of ${\cal G}(k^2)$ can be
sought from studies of the DSEs satisfied by the dressed gluon
propagator, $D_{\mu\nu}(k)$, and the dressed gluon-quark vertex
$\Gamma^a_\nu(q,p)$.  The latter is often represented by an Ansatz;
there is almost no information available from DSE studies; the gluon
propagator has been often studied via its DSE.  If the ghost loop and
the quark loop in the gluon DSE are unimportant, then the qualitative
conclusion from such studies is that the gluon propagator is
significantly enhanced in the infrared and well-represented by an
integrable singularity such as a regularization of $1/k^4$~\cite{BP89}.
Phenomenological studies containing such an enhancement show that
dynamical chiral symmetry breaking and quark confinement follow in a
straightforward and natural way from the quark DSE with an empirically
correct value for the chiral condensate $\langle \bar q q\rangle^0$ and
an excellent description of pion and kaon properties~\cite{MR97}.

Recent gluon DSE studies that include the ghost loop but not the quark
loop have suggested a weak infrared strength that vanishes at
\mbox{$k^2=0$} for the transverse component of $D_{\mu\nu}(k)$ due to a
strong infrared enhancement of the ghost propagator~\cite{AvSH98,AB98}.
In some studies of this type, unphysical particle-like singularities
occur in the Ansatz for the dressed ghost-gluon and 3-gluon
vertices~\cite{AvSH98}.  It is apparent that such gluon DSE studies are
presently limited by the type of truncation that can be accommodated and
the preliminary nature of the Ans\"atze employed for some of the dressed
vertices.  Several lattice studies of $D_{\mu\nu}(k)$ have been
interpreted in terms of an infrared behavior less singular than
$1/k^2$~\cite{LSW98}.  The phenomenological implications of either type
of non-singular infrared behavior for $D_{\mu\nu}(k)$ have recently been
explored within the quark DSE~\cite{HMR98}.  It was found that dynamical
chiral symmetry breaking as represented by a nonzero chiral condensate
is either absent or is a small fraction of what is required to explain
pion phenomena; the produced quark propagator does not show quark
confinement.

To provide a quark DSE-based description of pion and kaon phenomena as a
basis for exploring vector meson properties, we utilize a variation of
the following Ansatz introduced in Ref.~\cite{MR97}
\begin{equation}
\label{gk2}
\frac{{\cal G}(k^2)}{k^2} =
8\pi^4 D \delta^4(k) + \frac{4\pi^2}{\omega^6} D k^2 {\rm e}^{-k^2/\omega^2}
+ 4\pi\,\frac{ \gamma_m \pi}
        {\case{1}{2}
        \ln\left[\tau + \left(1 + k^2/\Lambda_{\rm QCD}^2\right)^2\right]}
{\cal F}(k^2) \,,
\end{equation}
with ${\cal F}(k^2)= [1 - \exp(-k^2/[4 m_t^2])]/k^2$, $\tau={\rm
e}^2-1$, and \mbox{$\gamma_m=12/(33-2N_f)$}.  This Ansatz preserves the
one-loop renormalization group behavior of QCD for solutions of the
quark DSE.  In particular, the correct one-loop QCD anomalous dimension
of the quark mass function $M(p^2)$ is preserved in its ultraviolet
behavior for both the chiral limit ($m(\mu)=0$, anomalous dimension
$1-\gamma_m$) and explicit chirally broken case ($m(\mu)=0$, anomalous
dimension $\gamma_m$).  This asymptotic behavior, a characteristic of
QCD, is confirmed by analysis of the numerical solution in the
ultraviolet as described in detail in Ref.~\cite{MR97}.  The main
qualitative feature of Eq.~(\ref{gk2}) is that the phenomenologically
required strong infrared enhancement in the region $0-0.5$~GeV$^2$ is
distributed over an integrable $\delta^4(k)$ singularity~\cite{mn83} and
a finite-width approximation to $\delta^4(k)$ normalized so that both
terms have the same $\int d^4k$.  The last term in Eq.~(\ref{gk2}) is
proportional to $\alpha(k^2)/k^2$ at large spacelike $k^2$ and has no
singularity on the real $k^2$ axis.  The parameters $\omega$ and $m_t$
were not varied freely in the study of Ref.~\cite{MR97}; the fixed
values $m_t=0.5$~GeV and $\omega=0.3$~GeV were chosen mainly to ensure
that ${\cal G}(k^2)\approx 4\pi\alpha(k^2)$ for $k^2>2\,$GeV$^2$.  The
free parameters were $D$ and the renormalized $u/d$- and $s$-quark
current masses to obtain a good description of $\pi$ and $K$ properties.

For the present study of vector mesons, we eliminate the
$\delta$-function term from Eq.~(\ref{gk2}) and allow the second
(finite-width) term to carry all of the infrared strength.  Solutions of
the rainbow DSE for the quark propagator, when investigated, usually
reveal a non-analytic behavior in the complex $p^2$-plane often in the
form of complex conjugate branch points~\cite{SC92,MH92} that are
modified or even eliminated when the gluon-quark vertex is
dressed~\cite{BRW92}.  Subsequent use of the propagator solutions in the
BSE for the bound state meson should be accompanied by a determination
that such non-analytic points (that are likely artifacts of the
truncation) lie outside the complex domain of integration that naturally
arises in the search for a solution of the BSE in Euclidean metric.  The
mass of the meson determines the extent of the required departures from
the quark real $p^2$ axis and the pion and kaon solutions from the
Ansatz of Eq.~(\ref{gk2}) are free of such problems.  However, with the
parameters of Ref.~\cite{MR97}, we have found this not to be the case
for the more massive vector solutions.  The removal of the
$\delta$-function term allows parameters to be easily found to preserve
the quality of the pion and kaon description while allowing numerically
accurate BSE solutions for the vector masses reported here.

We therefore employ the Ansatz
\begin{equation}
\label{gvk2}
\frac{{\cal G}(k^2)}{k^2} =
               \frac{4\pi^2}{\omega^6} D k^2 {\rm e}^{-k^2/\omega^2}
+ 4\pi\,\frac{ \gamma_m \pi}    {\case{1}{2}
        \ln\left[\tau + \left(1 + k^2/\Lambda_{\rm QCD}^2\right)^2\right]}
{\cal F}(k^2) \,.
\end{equation}
As in the earlier pion and kaon studies, we use \mbox{$m_t=0.5$~GeV},
\mbox{$\tau={\rm e}^2-1$}, \mbox{$N_f=4$}, \mbox{$\Lambda_{\rm
QCD}^{N_f=4}= 0.234\,{\rm GeV}$}, and a renormalization point
$\mu=19\,$GeV, which is sufficiently perturbative to allow the one-loop
asymptotic behavior of the quark propagator to be used as a check.  We
consider three parameter sets characterized by three different values of
$\omega$.  For each parameter set, $D$ is treated as a phenomenological
parameter, which was fitted, along with the renormalized current quark
masses, to obtain a good description of $m_{\pi/K}$ and $f_{\pi}$.
Subsequently, the vector meson sector was studied without parameter
adjustment.  For comparison we also report, for the Ansatz of
Ref.~\cite{MR97}, vector meson masses estimated by an extrapolation of
the BSE eigenvalue to the mass-shell point.

\section{Vector meson Bethe--Salpeter amplitudes}

The general form of a vector vertex $\Gamma_\mu(q;P)$ can be expressed
as a decomposition into twelve independent Lorentz covariants, made from
the three vectors $\gamma_\mu$, the relative momentum $q_\mu$, and the
meson total momentum $P_\mu$, each multiplied by one of the four
independent matrices $1\hspace{-3pt}$l, $\gamma\cdot q$, $\gamma\cdot
P$, and $\sigma_{\mu\nu} q_\mu P_\nu$.  Since a vector meson BS
amplitude is transverse the number of allowed covariants reduces to
eight, so that the general decomposition of the vector BS amplitude is
\begin{eqnarray}
\label{genvecbsa}
        \Gamma_\mu^V(q;P) & = & \sum_{i=1}^8 \, T^i_\mu(q;P) \, F_i(q^2,
                q\cdot P; P^2)\,,
\end{eqnarray}
with the invariant amplitudes $F_i(q^2, q \cdot P; P^2)$ being Lorentz
scalar functions.  The choice for the covariants $T^i_\mu(q;P)$ to be
used as a basis is constrained by the required properties under Lorentz
and parity transformations, but is not unique.  The BSE
Eq.~(\ref{matbse}) must be projected onto the covariant basis to produce
a coupled set of eight linear equations for the invariant amplitudes
$F_i$ to be cast in matrix form.  This requires a procedure to project
out a single amplitude from the general form Eq.~(\ref{genvecbsa}).  It
is therefore helpful if the chosen covariants satisfy a Dirac-trace
orthonormality property.

We have chosen the following set of dimensionless orthogonal covariants
\begin{eqnarray}
        T^1_\mu(q;P) & = & \gamma^T_\mu                         
\label{cov1}    \,,\\
        T^2_\mu(q;P) & = & \frac{6}{q^2\,\sqrt{5}}
                \left( q^T_\mu (\gamma^T\cdot q) - 
                \case{1}{3} \gamma^T_\mu (q^T)^2 \right)        \,,\\
        T^3_\mu(q;P) & = & \frac{2}{q\,P}
                \left( q^T_\mu (\gamma\cdot P) \right)          \,,\\
        T^4_\mu(q;P) & = & \frac{i\,\sqrt{2}}{q\,P}        
                \left(\gamma^T_\mu (\gamma\cdot P) (\gamma^T\cdot q) 
                + q^T_\mu (\gamma\cdot P) \right)               \,,\\
        T^5_\mu(q;P) & = & \frac{2}{q}\;q^T_\mu                 \,,\\
        T^6_\mu(q;P) & = & \frac{i}{q\,\sqrt{2}}
                \left( \gamma_\mu^T (\gamma^T\cdot q) - 
                (\gamma^T\cdot q) \gamma_\mu^T \right)          \,,\\ 
        T^7_\mu(q;P) & = & \frac{i\,\sqrt{3}}{q^2\,P\,\sqrt{5}}
        \left(1 - \cos^2{\theta}\right) \left(\gamma_\mu^T
        (\gamma\cdot P)-(\gamma\cdot P)\gamma_\mu^T\right)
        - \frac{1}{\sqrt{2}}\;T^8_\mu(q;P)                      \,,\\
        T^8_\mu(q;P) & = & \frac{i\,2\sqrt{6}}{q^2\,P\,\sqrt{5}}\;
                q^T_\mu (\gamma^T\cdot q)(\gamma\cdot P)        \,,
\label{cov8}
\end{eqnarray}
where $V^T$ is the component of $V$ transverse to $P$
\begin{eqnarray}
        V^T_\mu & = &  V_\mu - \frac{P_\mu\,(P\cdot V)}{P^2}\,,
\end{eqnarray}
and $q\cdot P = q\,P\,\cos\theta$.  Note that at the mass-shell
\mbox{$P = i\,m$}.  The orthonormality property satisfied by
these covariants is
\begin{equation}
\label{covnorm}
       \case{1}{12} {\rm Tr_D} \left[ 
        T^i_\mu(q;P) \,T^j_\mu(q;P) \right] 
        = f_i(\cos\theta) \delta_{ij}\,,
\end{equation}
where the functions $f_i(z)$ are given by $f_1(z) = 1$, 
$f_i(z) = \frac{4}{3}(1-z^2)$ for $i=3,4,5,6$ and 
$f_i(z) = \frac{8}{5}(1-z^2)^2$ for $i=2,7,8$.   For later use we also note 
the relation
\begin{equation}
\label{intnorm}
      \int_0^\pi\!d\theta\;\sin^2{\theta}  
        \; f_i(\cos\theta)= \case{\pi}{2} \,.
\end{equation}
The covariants are dimensionless and independent of the magnitudes $q$
and $P$.  These properties are helpful in allowing the relative
magnitude of the amplitudes $F_i$ to be a qualitative measure of the
dynamical importance of the various covariants.  A more quantitative
measure can depend on the particular observable being studied;
amplitudes that are unimportant at low momenta can become dominant when
high momentum behavior of the bound state solution is being probed.

For unflavored mesons that are eigenstates of $C$ (charge conjugation),
such as the $\rho$, $\omega$ and $\phi$, there is an additional
constraint on the BS amplitude\footnote{We do not discriminate between
up and down quarks, and do not take into account electromagnetic
corrections; therefore the BS amplitudes for $\rho^\pm$ are equal to
those for $\rho^0$.  Furthermore, the ladder truncation cannot
discriminate between isovector and isoscalar mesons; therefore the
$\rho$ and the $\omega$ are degenerate in this truncation.} to obtain a
specified $C$-parity.  Of the eight covariants given in
Eqs.~(\ref{cov1})-(\ref{cov8}), $T^3$ and $T^6$ are even under $C$, the
others are odd under $C$.  The only remaining quantity that can produce
a desired uniform $C$-parity is $q \cdot P$ which is odd under $C$.
Thus a $C=-$ solution (such as the $\rho$ and $\phi$) will have
amplitudes $F_3$ and $F_6$ that are odd in $q\cdot P$ while the
remaining amplitudes are even in $q\cdot P$.  For the flavored vector
meson $K^\star$, which is not an eigenstate of $C$, each amplitude will
contain both even and odd terms in $q\cdot P$.  Since the ladder
truncation of the BSE is invariant under charge conjugation if equal
momentum sharing ($\eta= 0.5$) is used, the observation of the above
odd-even behavior in $q\cdot P$ of $F_i$ can be used as a test of
numerical accuracy.  Alternatively, the amplitudes $F_i$ can be expanded
in terms of a basis of functions that are appropriately odd or even in
$\cos{\theta}$ to save significantly on computer time and memory.
Because the mass-shell condition makes the magnitude $P$ imaginary, it
is not difficult to verify that with definite $C$-parity, each amplitude
associated with our chosen basis of covariants is either purely real or
purely imaginary.  The amplitudes $F_i$ for the $K^\star$ solution are
in general complex due to the dependence upon all powers of $q\cdot P$.

After using the representation Eq.~(\ref{genvecbsa}) for the solution in
terms of the covariant basis, followed by projection using the
Dirac-trace orthonormality property Eq.~(\ref{covnorm}), the homogeneous
BSE Eq.~(\ref{matbse}) for a meson with flavored constituents $a\bar{b}$
reduces to a set of coupled integral equations for the eight functions
$F_i^{ab}(q^2,q\cdot P;P^2)$ in the form
\begin{eqnarray}
 F^{ab}_i(p^2,p\cdot P;P^2) f_i(z) &=& \case{4}{3}\int^\Lambda\!\!
        \frac{d^4q}{(2\pi)^4}\,{\cal G}((p-q)^2)\,
        D_{\mu\nu}^{\rm free}(p-q) F^{ab}_j(q^2,q\cdot P;P^2)
        \times \nonumber \\ & & \case{1}{12}{\rm Tr}_{D}\left[
        T_\rho^i(p;P) \gamma_\mu S^a(q+\eta P) T_\rho^j(q;P) 
        S^b(q-\bar\eta P) \gamma_\nu \right] \,.
\label{ampleqn}
\end{eqnarray}
The above system of equations was solved by two complementary methods.
The first method was a direct treatment as an integral eigenvalue
equation \mbox{$\lambda(P^2) {\sf F}={\sf K}(P^2) {\sf F}$} for a set of
functions ${\sf F}$ of two variables: $p^2$ and \mbox{$z=\cos\theta$}.
An iterative method is used to determine the smallest $m$ satisfying
\mbox{$\lambda(-m^2)=1$}.  Both variables were discretized via Gaussian
quadrature and the summations for the double integration were carried
out at each iteration.  This has a high demand on computer memory.

In the second method, the angle dependence of the amplitudes is expanded
in the form
\begin{eqnarray}
\label{chebexp}
        F_i(q^2,q\cdot P;P^2) &=& \sum_{j=0}^{\infty}
                \,{}^j\!F_i(q^2;P^2)\,U_j(\cos\theta)\,,
\end{eqnarray}
where the $U_j(z)$ are Chebyshev polynomials of the second kind.  This
allows the angle integrations in Eq.~(\ref{ampleqn}) to be carried out
to produce an integral equation in one variable but for a larger set of
functions ${}^j\!F_i(q^2;P^2)$.  For $C=-$ eigenstates such as $\rho$
and $\phi$, amplitudes $F_3$ and $F_6$ will require only odd order
Chebyshev terms while the other amplitudes will require only even terms.
In practice, the number of Chebyshev terms required is quite low (one or
two terms) so that the memory requirements are effectively reduced in
this second method.  The solutions from the direct two-variable approach
can be projected onto the Chebyshev basis as a check on the second
method and also as a means of presentation.

The specific normalization condition for the vector meson solutions of
the ladder BSE follows from Eq.~(\ref{gennorm}) and is
\begin{eqnarray}
\label{vecnorm}
    \left. 2 P_\mu = \frac{\partial}{\partial P_\mu} \,
        \frac{N_c}{3} \int^\Lambda_q {\rm Tr}_{D}
        \left[\bar\Gamma_\nu^{ba}(q;-K)\, S^a(q+\eta P)\, 
        \Gamma_\nu^{ab}(q;K)\,S^b(q-\bar\eta P)\right]
        \right|_{P^2=K^2=-m^2} \,,
\end{eqnarray}
where the factor $1/3$ appears because the three transverse
directions are summed.

\section{Electroweak decay}

Here we summarize the definition of, and our convention for, the vector meson
leptonic and electromagnetic decay constants and their explicit relationship 
to the BS amplitudes.  The electromagnetic decay mediated by a photon
(e.g. $\rho^0$, $\omega$, $\phi$), and the leptonic decay mediated by a 
W-boson (e.g. $\rho^\pm$, $K^{\star\pm}$), are described by the vector decay 
constant defined by~\cite{GasLeut}
\begin{eqnarray}
\label{fv}
   f_V\, m_V\, \epsilon_\mu^{(\lambda)}(P) &=& 
        \langle 0|\bar q^b \gamma_\mu q^a| V^{ab}(P,\lambda)\rangle \,,
\end{eqnarray}
where $\epsilon_\mu^{(\lambda)}$ is the polarization vector of the vector 
meson satisfying \mbox{$\epsilon^{(\lambda)}\cdot P = 0$}
and normalized such that 
\mbox{$\epsilon_\mu^{(\lambda)*}\,\epsilon_\mu^{(\lambda)}=3$}.  This is 
completely analogous to the definition
\begin{eqnarray}
   f_P\, P_\mu\, &=& 
        \langle 0|\bar q^b \gamma_\mu \gamma_5 q^a| P^{ab}(P)\rangle
\end{eqnarray}
for the pseudoscalar decay constant that corresponds to $f_\pi =
131$~MeV under the normalization convention of Eq.~(\ref{gennorm}).  The
vector decay constant from Eq.~(\ref{fv}) can be expressed as the
loop-integral
\begin{eqnarray}
\label{decayconst}
   f_V m_V &=& \frac{Z_2\,N_c}{3} 
        \int^\Lambda\!\!\frac{d^4q}{(2\pi)^4}\,
        {\rm Tr}_{D}\left[\gamma_\mu S^a(q+\eta P) 
        \Gamma^{ab}_\mu(q;P) S^b(q-\bar{\eta}P)\right]\,,
\end{eqnarray}
which is exact if the dressed quark propagators and the meson BS
amplitude are exact~\cite{IKR99}.  In the next section we use
Eq.~(\ref{decayconst}) to calculate the decay constants $f_\rho$,
$f_\phi$ and $f_{K^\star}$.

The coupling of the $\rho^0$
\begin{eqnarray}
\label{rhozero}
        |\rho^0\rangle &=& \frac{1}{\sqrt{2}}
                \left(|u \bar u\rangle -  |d \bar d\rangle \right) 
\end{eqnarray}
to the photon is conventionally expressed via a dimensionless coupling
constant $g_\rho$ in the form
\begin{eqnarray}
\label{grho}
   \frac{m_\rho^2}{g_\rho}\; \epsilon_\mu^{(\lambda)}(P) &=& 
        \langle 0|\bar{\cal Q} \, \hat{\rm Q} \,
        \gamma_\mu {\cal Q}| \rho^0(P,\lambda)\rangle\,,
\end{eqnarray}
where the flavor multiplet of quark field spinors is ${\cal Q} = {\rm
column}(u,d)$, and $\hat{\rm Q}$ is the quark electromagnetic charge
operator.  The normalization condition given in Eq.~(\ref{vecnorm}) is
in a form appropriate for a single flavor configuration $\bar{q_a}q_b$,
not for a multi-flavor configuration state like the $\rho^0$.  For such
states, Eq.~(\ref{vecnorm}) can be generalized by promoting the quark
propagators to flavor matrices ${\cal S} = {\rm diag}(S^u, S^d)$,
multiplying BS amplitudes $\Gamma_\mu$ by the appropriate flavor matrix,
and tracing over flavor indices as well.  The isospin-symmetric limit
with $S^u = S^d$ produces BS amplitudes that are independent of flavor
labels; the $\rho^0$ BS amplitude, for example, can then be expressed as
$(\tau_3/\sqrt{2})\,\Gamma_\mu$ where $\Gamma_\mu$ is the normalized BS
amplitude for the $\rho^\pm$.  Use of Eq.~(\ref{rhozero}) in
Eq.~(\ref{grho}) then gives
\begin{eqnarray}
   \frac{m_\rho^2}{g_\rho} &=& \frac{Z_2\,N_c}{3\,\sqrt{2}}\,
        \int^\Lambda\!\!\frac{d^4q}{(2\pi)^4}\, 
        {\rm Tr}_{D}\left[\gamma_\mu S^{u=d}(q+\eta P) 
        \Gamma^{ab}_\mu(q;P) S^{u=d}(q-\bar{\eta}P)\right] 
\nonumber \\ &=& \frac{f_\rho m_\rho}{\sqrt{2}}\,.  
\label{rhophoton}
\end{eqnarray}

The decay width $\Gamma_{\rho^0\rightarrow e^+e^-} =
6.77$~keV~\cite{PDG} leads via
\begin{eqnarray}
        \Gamma_{\rho^0 \rightarrow e^+\,e^-} &=& 
                \frac{4\pi\,\alpha^2\,m_\rho}{3\;g_\rho^2}
\end{eqnarray}
to the value \mbox{$g_\rho =5.03$}, that is \mbox{$f_\rho = 216$}~MeV.
Note that the isoscalar version of these considerations produces an
extra factor of $1/3$ on the right of Eq.~(\ref{rhophoton}) for the
coupling of the $\omega$ to a photon.  The partial width $\Gamma_{\omega
\rightarrow e^+e^-}$ is indeed about 10 times smaller than
$\Gamma_{\rho^0\rightarrow e^+e^-}$.

In a similar way, the coupling of the photon to the $\phi$, assumed to
be a pure $s\bar s$-state, is defined as
\begin{eqnarray}
   \frac{m_\phi^2}{g_\phi}\; \epsilon_\mu^{(\lambda)}(P) &=& 
        {\textstyle\frac{1}{3}}\,
        \langle 0|\bar s \gamma_\mu s| \phi(P,\lambda)\rangle\,,
\end{eqnarray}
and the relation between $g_\phi$ and the vector decay constant
$f_\phi$ is
\begin{eqnarray}
   \frac{m_\phi^2}{g_\phi} &=&   \frac{ f_\phi m_\phi} {3}      \\
        &=& \frac{Z_2\,N_c}{9}
        \int^\Lambda\!\!\frac{d^4q}{(2\pi)^4}\, 
        {\rm Tr}_{D}\left[\gamma_\mu  S^{s}(q+\eta P) 
        \Gamma^{ss}_\mu S^{s}(q-\bar{\eta}P)\right]\,.
\end{eqnarray}
The partial width of the $\phi \rightarrow e^+e^-$ decay is
\begin{eqnarray}
        \Gamma_{\phi \rightarrow e^+\,e^-} &=& 
                \frac{4\pi\,\alpha^2\,m_\phi}{3\;g_\phi^2}\,,            
\end{eqnarray}
and the experimental value $1.37 \pm 0.05$~keV~\cite{PDG} produces
$f_\phi = 237$~MeV, that is $g_\phi = 12.9$.

The decay constant $f_V$ determines not only the coupling of the neutral
vector mesons to a photon, but also the coupling of $\rho^\pm$ and
$K^{\star\,\pm}$ to the weak vector bosons W$^\pm$.  There are no data
available for the leptonic decay of these charged vector mesons, but the
couplings can be extracted indirectly from the decays 
\mbox{$\tau \rightarrow \rho\,\nu_\tau$} and 
\mbox{$\tau \rightarrow K^\star\,\nu_\tau$}.  The partial width for 
such a decay is
\begin{eqnarray}
\label{taudecay}
  \Gamma_{\tau \rightarrow V\nu_\tau} &=&
        \frac{G_F^2\,m_\tau}{8\,\pi} \,V_{ab}^2\,f_V^2\,m_V^2
        \left(1-\frac{m_V^2}{m_\tau^2}\right)^2
        \left(1+\frac{m_\tau^2}{2\,m_V^2}\right) \,.
\end{eqnarray}
With the experimental values for the partial decay width~\cite{PDG}
\mbox{$\Gamma_{\tau \rightarrow \rho\,\nu_\tau} = 25.02\,\%\;
\Gamma_{\hbox{\scriptsize total}}$} and \mbox{$\Gamma_{\tau \rightarrow
K^\star\,\nu_\tau} = 1.28\,\%\;\Gamma_{\hbox{\scriptsize total}}$}, and
the CKM matrix elements $V_{ud}=0.974$ and $V_{us}=0.220$, this gives a
ratio
\begin{eqnarray}
\label{fkstar}
        \frac{f_{K^\star}}{f_\rho} = 1.042
\end{eqnarray}
and thus a decay constant $f_{K^\star} = 225$~MeV, if we use the
experimental value $f_\rho = 216$~MeV.

With the available data, the absolute value of $f_\rho$ using
Eq.~(\ref{taudecay}) gives $f_\rho = 208$~MeV.  We expect however that
the direct determination of $f_\rho$ through $\rho^0\rightarrow e^+e^-$,
giving $f_\rho = 216$~MeV, is a more accurate determination of this
decay constant.  In particular, most higher-order corrections to the
electroweak vertex are likely to cancel in the ratio of the partial
decay widths, and therefore we use the ratio in Eq.~(\ref{fkstar}) to
extract the experimental $f_{K^\star}$.

\section{Numerical results}

\begin{figure}
\centering{\
\epsfig{figure=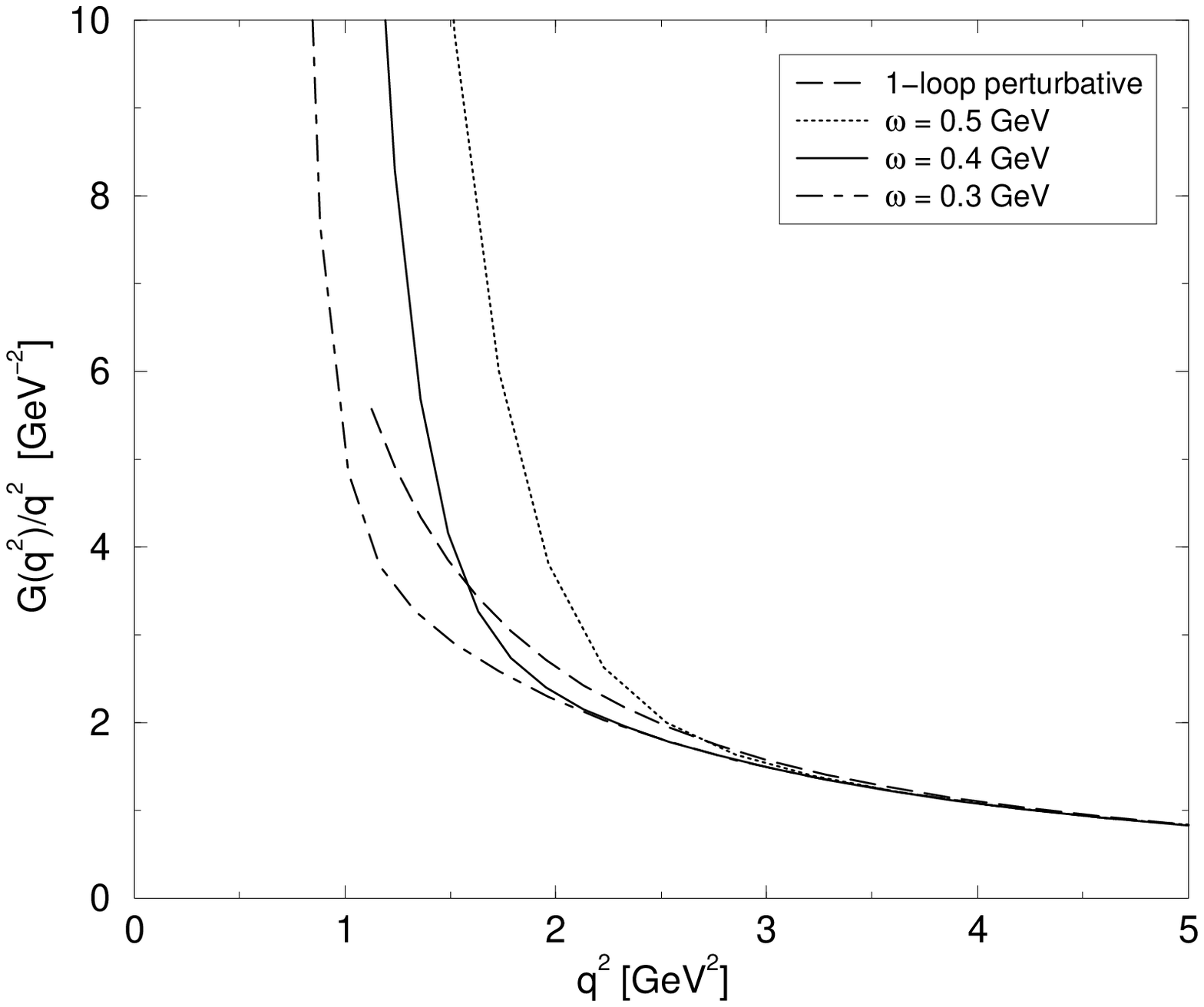,height=9.0cm}}
\caption{
The Ansatz for the effective $\bar q q$ interaction ${\cal G}(q^2)/q^2$,
Eq.~(\ref{gvk2}), for the three parameter sets, together with the 1-loop
perturbative result for comparison.
\label{figgluon}}
\end{figure}

In Fig.~\ref{figgluon} we show our Ansatz for the effective interaction,
Eq.~(\ref{gvk2}), for the three different parameter sets we have
explored, characterized by the values of $\omega$, together with the
1-loop perturbative coupling for comparison.  We use three different
values of the parameter $\omega$, constrained only by the requirement
that the perturbative coupling above $q^2 = 3\ {\rm GeV}^2$ should be
reproduced.  It is only in the infrared region, below $q^2 = 2\ {\rm
GeV}^2$, that there is a significant difference between the three
parameterizations and the perturbative result.  The parameter $D$ and the
current quark mass $m_{u/d}(\mu)$ are fixed by fitting $m_\pi$ and
$f_\pi$.  Next, the strange quark mass $m_{s}(\mu)$ is determined by a
fit to the kaon mass.  The resulting value of the kaon decay constant
$f_K$ is within 3\% percent of the experimental value, almost
independent of the parameter set for the effective interaction.  All
three parameter sets lead to a good description of the pion and kaon
masses and decay constants, as well as a reasonable value of the chiral
condensate.  In Table~\ref{respseudo} we have summarized these results
for the three different parameter sets, together with the results from
Ref.~\cite{MR97}.  

With our parameterization, the quark mass function $M(p^2)=B(p^2)/A(p^2)$
has qualitatively the same behavior as obtained in Ref.~\cite{MR97}.
With a Euclidean constituent-quark mass $M^E$ defined as the solution of
$p^2 = M^2(p^2)$, we obtain constituent quark masses of about
\mbox{$M_{u/d} = 300 - 500\ {\rm MeV}$} for the light quarks, and $M_s =
500 - 640\ {\rm MeV}$ for a strange quark, spanned by the three
parameter sets; the parameterization of Ref.~\cite{MR97} gives
constituent masses \mbox{$M_{u/d} = 560\ {\rm MeV}$} and $M_s = 700\
{\rm MeV}$.

\subsection{Results for vector meson observables}

In Table~\ref{resvecvarious} we present our results for the vector meson
masses and decay constants.  The full angular dependence was retained in
the calculation of these results: we solve the set of integral equations
Eq.~(\ref{ampleqn}) with the $F_i(p^2,p\cdot P;P^2)$ treated as
functions of two variables $p^2$ and $z=\cos\theta$.  This eigenvalue
problem defines physical solutions at the mass-shell $P^2 = -m_V^2$.
All calculations with the gluon Ansatz of Eq.~(\ref{gvk2}) were
performed at the physical mass-shell; the calculations we have performed
with the parameterization of Ref.~\cite{MR97} for comparison involved
some extrapolation\footnote{The extrapolations were necessary because of
nonanalytic behavior of the resulting quark propagator as discussed in
Sec.~\ref{secladder}.} to the mass-shell, which makes these results less
accurate.  In particular the integral for the normalization condition,
Eq.~(\ref{vecnorm}), is very sensitive to such an extrapolation, which
is why we do not report the decay constant for this particular model.

All parameterizations we used give equally good results for the masses
and decay constants: the results are fairly insensitive to changes in
$\omega$ and $D$, as long as they are fit to $m_\pi$, $f_\pi$ and $m_K$.
Our result for $m_\rho$ is typically 5\% too low, whereas $m_{K^\star}$
and $m_\phi$ are typically 5\% too large.  Our result for the decay
constants are within 10\% of the experimental value for $f_\rho$ and
$f_{K^\star}$, and within 10\% to 15\% for $f_\phi$, depending on the
parameter set.  This agreement with experiment is quite encouraging,
given the fact that the parameters are fixed by pseudoscalar
observables.

From Table~\ref{resvecvarious} we can also conclude that only five of
the eight covariants are qualitatively and quantitatively important for
the vector meson masses and decay constants; this seems to be general,
i.e. independent of the parameter set used.  Of course, the relative
importance of different covariants in a BS amplitude does depend on the
observable under consideration. Also, use of a basis set of eight
independent covariants that is different from the present basis given in
Eqs.~(\ref{cov1})-(\ref{cov8}), could produce a different conclusion
concerning the number of important covariants.

\begin{figure}
\centering{\
\epsfig{figure=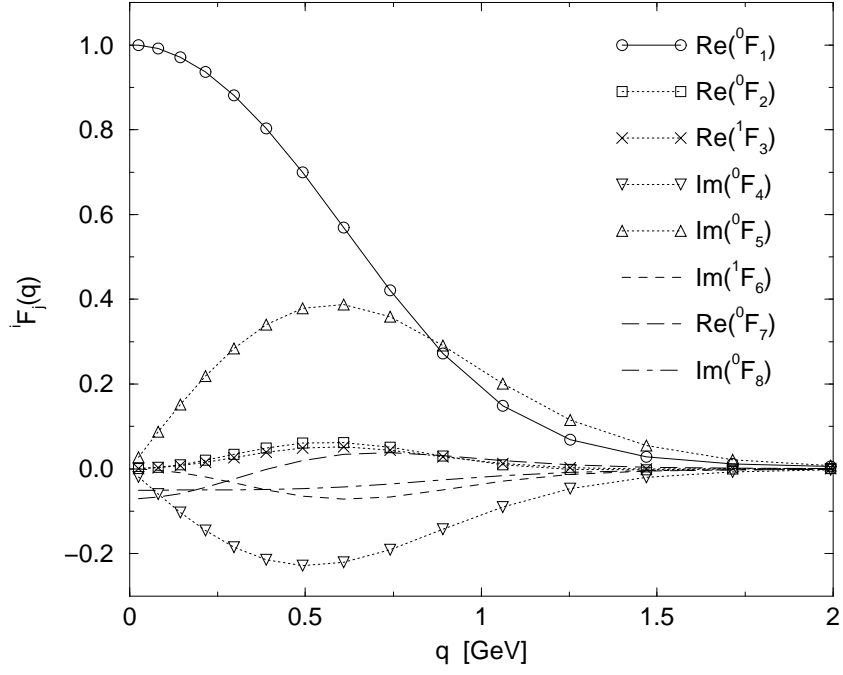,height=9.0cm}}
\caption{
The leading Chebyshev projections of all eight $\rho$ BS amplitudes,
normalized to ${}^0\!F_1(0) = 1$, with an effective $\bar q q$
interaction, Eq.~(\ref{gvk2}), with $\omega = 0.4\,{\rm GeV}$, $D =
0.93\,{\rm GeV}^2$.  The most important amplitudes, $F_1$-$F_5$, are
labeled by lines with symbols.
\label{figrhoBSA}}
\end{figure}

In Fig.~\ref{figrhoBSA} we show the behavior of the leading Chebyshev
projection of the invariant amplitudes of the $\rho$ BS amplitude,
${}^0\!F_i^\rho(q^2;P^2)$.  This and the other plots of the BS
amplitudes are produced with the parameter set 
\mbox{$\omega = 0.4\ {\rm GeV}$} and $D = 0.93\ {\rm GeV}^2$; 
the results for the other parameter sets look qualitatively the same.
The leading amplitude for the pion, $E_\pi$, and for the rho, $F_1$, are
very similar; however, this similarity might be accidental.  Of the
sub-dominant amplitudes, $F_4$ and $F_5$ are significantly larger than
the rest.  The magnitude of the amplitudes $F_6$, $F_7$, and $F_8$ is
much smaller than that of $F_1$, $F_4$, and $F_5$, as is evident from
Fig.~\ref{figrhoBSA}; this makes it understandable why these amplitudes
contribute so little to the vector meson masses and decay constants.
From this figure one might conclude that the amplitudes $F_2$ and $F_3$
have a similar minor role.  However, it turns out that these amplitudes
are essential for the convergence of the loop integral for the decay
constant, Eq.~(\ref{decayconst}), as discussed below, in
Sec.~\ref{secBSUVas}.

To study the relevance of the various covariants for physical
observables in more detail, we calculate the vector meson masses and
decay constants using different subsets of the eight covariants in our
basis.  These results are given in Table~\ref{resvecnew} for one
particular parameter set, together with the results from use of only the
leading Chebyshev moments of each amplitude $F_i(q^2,q\cdot P;P^2)$.
Note that the leading Chebyshev order for the $K^\star$ is zeroth order
for all amplitudes $F_i$, in contrast to the case for the $\rho$ and
$\phi$: the functions $F_3$ and $F_6$ are odd in $q\cdot P$ for the
$\rho$ and $\phi$ because of charge conjugation symmetry, so the leading
Chebyshev order is $U_1(\cos{\theta})$ for those mesons.  It is evident
that for the $\rho$ and $\phi$ only the leading Chebyshev moment is
needed to get accurate results for the masses and decay constants; but
the second Chebyshev moment of $F_1$ is needed for strict convergence of
Eq.~(\ref{decayconst}).  We expect this to be a general phenomenon:
practical calculations of hadron observables might be facilitated by a
suitable parameterization of the leading Chebyshev moments of the
amplitudes $F_1$ through $F_5$.  For the $K^\star$, which is not a
charge conjugation eigenstate, one needs at least the zeroth and the
first Chebyshev moments for an accurate description.

Another difference between the $K^\star$ and the $\rho$ and $\phi$
mesons, is the dependence on the momentum sharing parameter $\eta$ in
Eq.~(\ref{ampleqn}).  Charge conjugation dictates use of $\eta = 0.5$
for the $\rho$ and the $\phi$.  For the $K^\star$ there is no such
constraint and we explored momentum partition sets
$(\eta_{u},\bar\eta_s)$ varying between $(0.5,0.5)$ and $(0.4,0.6)$.
Physical observables are in principle independent of this partitioning;
any dependence of $K^\star$ physical observables on
$(\eta_{u},\bar\eta_s)$ would signal an inadequacy of the ladder
truncation or subsequent approximations.  We find that the results for
$m_{K^\star}$ and $f_{K^\star}$ are indeed unchanged under variation of
the momentum sharing, {\em as long as all covariants and the full
angular dependence are taken into account}.  Once certain amplitudes are
dropped and/or the angular dependence of the amplitudes is truncated,
physical observables do become dependent on $(\eta_{u},\bar\eta_s)$:
variations between $(0.5,0.5)$ and $(0.4,0.6)$ lead to changes in
$m_{K^\star}$ and $f_{K^\star}$ of up to 5\%.

\begin{figure}
\centering{\
\epsfig{figure=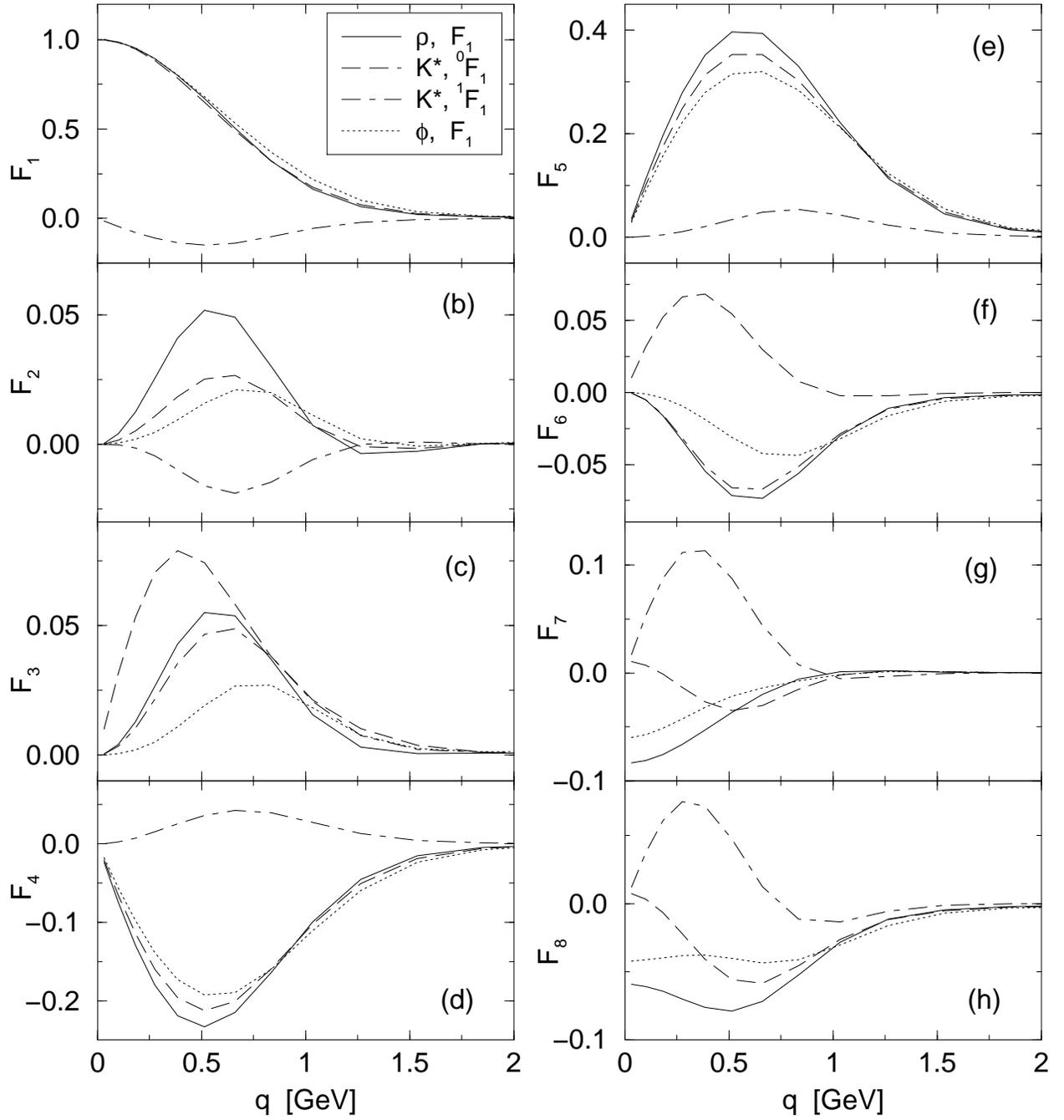,height=18.0cm}}
\caption{
Leading and subleading BS amplitudes for the $\rho$, $K^\star$, and
$\phi$ mesons, (a) zeroth Chebyshev projections of $F_1$, and for the
$K^\star$ also the first Chebyshev projection, (b) as (a), but then for
$F_2$, (c) first Chebyshev projections of $F_3$, and for the $K^\star$
also the zeroth projection, (d) as (a), but then for $F_4$, (e) as (a),
but then for $F_5$, (f) as (c), but then for $F_6$, (g) as (a), but then
for $F_7$, (h) as (a), but then for $F_8$. The parameters are the same
as in the previous plot: $\omega=0.4\,{\rm GeV}$, $D=0.93\,{\rm GeV}^2$.
\label{figvecBSA}}
\end{figure}

A comparison of the BS amplitudes of the three different vector mesons
is made in Fig.~\ref{figvecBSA}.  This figure clearly shows the
difference between the $\rho$ and $\phi$ mesons on the one hand, and the
$K^\star$ on the other: while the leading Chebyshev moments of the
$\rho$ and $\phi$ amplitudes are very similar to each other and to the
corresponding moments of the $K^\star$ amplitude, the latter has both
even and odd moments, due to the lack of C-parity.  This is especially
evident for the amplitudes $F_3(q,q\cdot P;P^2)$ and $F_6(q,q\cdot
P;P^2)$, which have no zeroth Chebyshev moment in the case of the $\rho$
and $\phi$, but have a significant zeroth Chebyshev moment for the
$K^\star$.

\subsection{Asymptotic behavior of the BS amplitudes}
\label{secBSUVas}

The asymptotic behavior of the BS amplitudes for the $\rho$-meson is
shown in Fig.~\ref{figrhoUV}, and as in the pseudoscalar case, all
amplitudes behave like $1/q^2$ or $1/q^3$, up to calculable logarithmic
corrections.  We emphasize that in QCD these logarithmic corrections are
essential for the convergence of the integral for the decay constant.
Evaluation of the trace in Eq.~(\ref{decayconst}) for equivalent flavors
and equal momentum partitioning gives the leading behavior
\begin{eqnarray}
\label{decayUV}
f_V m_V &=& \frac{Z_2\,N_c}{3} \int^\Lambda 
        \frac{d^4q}{(2\pi)^4}\Big\{
        \left(12\sigma_s^+\sigma_s^- + 
        (4 q^2 + 8 q^2 \cos^2{\theta} - 3P^2)\sigma_v^
        +\sigma_v^- \right) \,F_1(q^2,q\cdot P;P^2)
\nonumber \\ 
        & - & 32\,q^2 \left(1 - \cos^2{\theta} \right)^2
        \sigma_v^+\sigma_v^- \,F_2(\ldots) /\sqrt{5}
        - 16\,q^2\,\cos{\theta}\,\left(1 - \cos^2{\theta} \right) 
        \sigma_v^+\sigma_v^-\,F_3(\ldots)
\nonumber \\  
        & + & i 8 \sqrt{2}\, P q \left(1 - \cos^2{\theta} \right) 
        \sigma_v^+\sigma_v^-\,F_4(\ldots)
        - i 8\, q \left(1 - \cos^2{\theta} \right) 
        \left(\sigma_v^+\sigma_s^- + \sigma_s^+\sigma_v^-\right) 
        \,F_5(\ldots)
\nonumber \\
     &+&  {\cal O}\left(\left(F_6 + F_7 + F_8\right) 
                        q\,\sigma_v^\pm \sigma_s^\mp\right) \Big\} \,,
\end{eqnarray}
where $\sigma_{v,s}$ are the vector and scalar components of the quark
propagator
\begin{eqnarray}
 \sigma_v&=&\frac{A(q)}{A^2(q)\,q^2 + B^2(q)} \,,\\
 \sigma_s&=&\frac{B(q)}{A^2(q)\,q^2 + B^2(q)} \,,
\end{eqnarray}
and $f^\pm := f(q_\pm)$.  The last terms in Eq.~(\ref{decayUV}),
proportional to $F_6$, $F_7$, and $F_8$, give small and convergent
contributions, since they behave in the ultraviolet as
\begin{equation}
     F_i(q^2,q\cdot P;P^2) q\,\sigma_v^\pm \sigma_s^\mp
        \;\sim\; \frac{1}{q^5} \,,
\end{equation}
up to logarithmic corrections.  Both contributions involving $F_4$ and
$F_5$, which fall off like $1/q^3$, are also ultraviolet finite.  Since
the amplitudes $F_1$, $F_2$, and $F_3$ fall off as $1/q^2$, simple power
counting shows that their individual contributions to
Eq.~(\ref{decayUV}) are logarithmically divergent, even accounting for
the cut-off dependence of $Z_2(\Lambda^2, \mu^2)$\footnote{The factor
$Z_2(\Lambda^2,\mu^2)$ ensures gauge invariance and cancels logarithmic
divergences in covariant gauges other than Landau gauge.}.

In order to analyze the asymptotic behavior produced by the vector meson
BSE in more detail, we follow the strategy used in Ref.~\cite{atkjohn}
for the asymptotic behavior of the function $B(p^2)$ from the quark DSE.
The key step is to replace the effective running coupling ${\cal
G}((p-q)^2)$ by ${\cal G}(\max{(p^2,q^2)})$.  In the ultraviolet region
the running coupling behaves like $1/\ln(y)$ with $y = k^2 /
\Lambda_{\rm QCD}^2$ which is a slowly varying function; therefore the
error made in using this approximation is under control.  In the
infrared region, such an approximation is not to be trusted.  After this
approximation, and with use of the Chebyshev decomposition for the
angular dependence of $F_i(q^2, q\cdot P,P^2)$ in Eq.~(\ref{ampleqn}),
all angular integrations can be performed analytically.  For the leading
Chebyshev moments of the $\rho$ BS amplitudes, Eq.~(\ref{ampleqn})
produces integral equations of the form
\begin{eqnarray}
\label{genUVinteq}
  F_i(x) &=& \frac{\gamma_m}{\ln{x}} 
        \int_0^x dy\; {\rm K}_{x>y}^{ij}(x,y)\,F_j(y) 
        + \gamma_m \int_x^\infty dy\; {\rm K}_{y>x}^{ij}(x,y) 
        \frac{F_j(y)}{\ln{y}} \,,
\end{eqnarray}
where $x = p^2/\Lambda_{\rm QCD}^2$, $y = q^2/\Lambda_{\rm QCD}^2$,
$F_i(x) = {}{}^0\!F_i(q^2;P^2)$ for $i=1,2,4,5,7,8$ and $F_i(x) =
{}{}^1\!F_i(q^2;P^2)$ for $i = 3,6$.  Now the coupled integral equations
can be converted to a set of coupled linear differential equations,
which can be solved in the ultraviolet region by assuming a series
expansion in both $x$ and $\ln{x}$.  For completeness, we have given the
relevant kernels ${\rm K}^{ij}$ and other details in
Appendix~\ref{AppAsRho}.

The analysis in Appendix~\ref{AppAsRho} shows that the ultraviolet
behavior of the amplitudes $F_1(x)$, $F_2(x)$ and $F_3(x)$, is of the
form
\begin{eqnarray}
        F_i(x) &=& \frac{a_i\,(\ln{x})^\alpha }{x}
       \left( 1 + \sum_{j=1}^\infty c_j\,(\ln{x})^{-j} \right) \,,
\end{eqnarray}
and the steps leading to identification of the power $\alpha$ and the
leading coefficients $a_i$ are also given there.  The leading
ultraviolet behavior is found to be
\begin{eqnarray}
\label{UVBSA1}
 {}^0\!F_1(q^2;P^2) \;=\; F_1(x) &\sim& 
        \frac{a_1\,(\ln{x})^\alpha}{x}                  \,, \\
\label{UVBSA2}
 {}^0\!F_2(q^2;P^2) \;=\; F_2(x) &\sim& 
        \frac{a_1\;2\,\sqrt{5}\,(\ln{x})^\alpha}{9\,x}  \,,\\
\label{UVBSA3}
 {}^1\!F_3(q^2;P^2) \;=\; F_3(x) &\sim& 
        \frac{a_1\,(\ln{x})^\alpha}{3\,x}               \,,
\end{eqnarray}
with 
\begin{eqnarray}
\label{UValpha}
        \alpha &=& -1 + \gamma_m/108                    \,.
\end{eqnarray}
The overall constant $a_1$ is not determined by the homogeneous BSE; its
value follows from the normalization condition.

\begin{figure}
\centering{\
\epsfig{figure=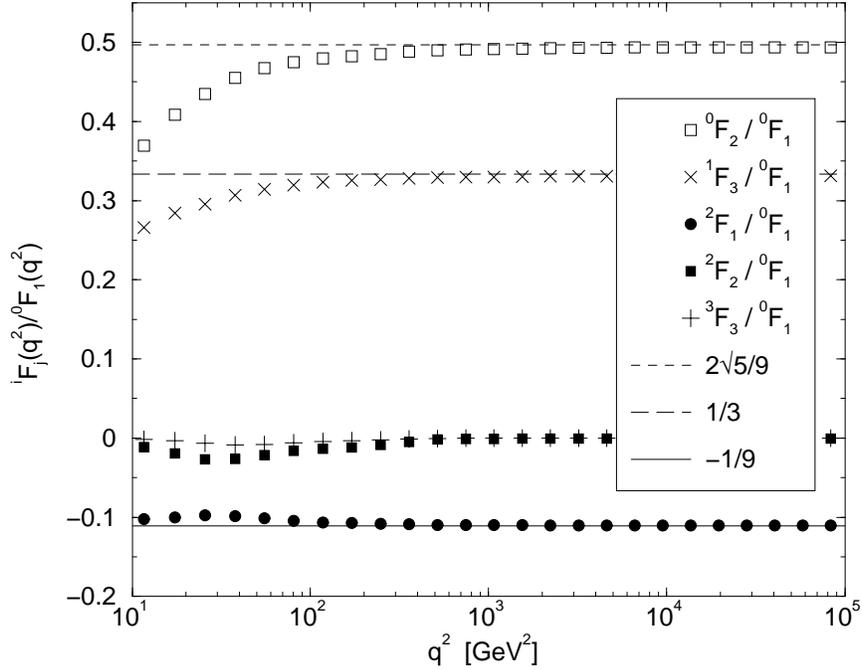,height=9.0cm}}
\caption{
The ultraviolet behavior of the $\rho$ BS amplitudes: the ratio
${}^k\!F_i(q^2;P^2)/{}^0\!F_1(q^2;P^2)$ for the leading amplitudes.
\label{figrhoUVratio}}
\end{figure}

Our numerical results show that the leading ultraviolet behavior of the
BS amplitudes is governed not only by the leading Chebyshev moments
${}^0\!F_1(q^2;P^2)$, ${}^0\!F_2(q^2;P^2)$, and ${}^1\!F_3(q^2;P^2)$,
but also by the second Chebyshev moment ${}^2\!F_1(q^2;P^2)$, see
Fig.~\ref{figrhoUVratio}.  Numerically, we find in the ultraviolet
\begin{eqnarray}
\label{numUVBSA2}
 \frac{{}^0\!F_2(q^2;P^2)}{{}^0\!F_1(q^2;P^2)} &=& 0.48 \pm 0.01 \,, \\
\label{numUVBSA3}
 \frac{{}^1\!F_3(q^2;P^2)}{{}^0\!F_1(q^2;P^2)} &=& 0.33 \pm 0.01 \,,  \\
\label{numUVBSA21} 
 \frac{{}^2\!F_1(q^2;P^2)}{{}^0\!F_1(q^2;P^2)} &=& -0.11 \pm 0.005 \,,
\end{eqnarray}
while all other ${}^k\!F_i$ fall off faster.  This is in excellent
agreement with the analytical results for the relative magnitudes of the
leading Chebyshev components, Eqs.~(\ref{UVBSA1})-(\ref{UVBSA3}).  The
power $\alpha$ of the logarithm is much harder to determine numerically;
our results indicate $-0.95 < \alpha < -1.0$, which is consistent with
$\alpha = -0.996$ from Eq.~(\ref{UValpha}).  We have not studied whether
the inclusion of ${}^2\!F_1$ in the analysis of the asymptotic behavior
would change our analytical result for $\alpha$; our numerical results
indicate that it will not influence the coefficients $a_i$ of
${}^0\!F_1$, ${}^0\!F_2$, and ${}^1\!F_3$, nor will it change the power
$\beta$ in Eq.~(\ref{BSasym}).

The ultraviolet behavior of the integral for the decay constant,
Eq.~(\ref{decayUV}), can now be analyzed in more detail.  The
ultraviolet behavior of the functions $F_1$, $F_2$, and $F_3$ does
indeed lead to individual divergent integrals for $\alpha \geq -1$.
However, the {\em combined} contribution is
\begin{eqnarray}
 f_V m_V &\sim& \int^{\Lambda^2} dy\; \frac{(\ln{y})^\alpha}{y} 
     \; \left\{ 6\,a_1 - 4\sqrt{5}\,a_2 - 4\,a_3 + 2\,b_1 \right\} \,,
\end{eqnarray}
where $b_1$ is the coefficient of the second Chebyshev moment
${}^2\!F_1(q^2;P^2)$, that is, the counterpart of $a_1$ in
Eq.~(\ref{UVBSA1}).  Use of the asymptotic behavior we have found
analytically, $a_2 = a_1\,2\,\sqrt{5}/9$ and $a_3 = a_1/3$, shows that
the integral for the decay constant is finite if $b_1 = -a_1/9$, which
agrees with our numerical result, Eq.~(\ref{numUVBSA21}).  This
cancellation between naive divergences coming from different covariants
provides an illustration of how renormalizibility is realized; it is
expected since the one-loop renormalization group behavior of QCD is
preserved in our rainbow-ladder truncation of the DSE and BSE.  It is
the vector counterpart of a similar cancellation in the integral for the
pseudoscalar decay constant: numerically~\cite{MR97}, it was found that
in the ultraviolet region the $\pi$ BS amplitudes satisfy $G_\pi = 2
F_\pi/q^2$, which makes the integral for $f_\pi$ finite, although the
separate contributions from $F_\pi$ and $G_\pi$ diverge.  The above
analysis, when applied to the pseudoscalar BSE, produces an asymptotic
behavior of the amplitudes $F_\pi$ and $G_\pi$ that exactly gives $G_\pi
= 2 F_\pi/p^2$.

Finally, in Appendix~\ref{AppAsRho}, the same ultraviolet analysis
applied to the vector BS amplitudes $F_4$ and $F_5$ gives
\begin{eqnarray}
\label{UVBSA4}
 F_4(x) &\sim& \frac{a_4}{x^{3/2}\,(\ln{x})^{1-\frac{1}{3}\gamma_m}}\,, \\
\label{UVBSA5}
 F_5(x) &\sim& \frac{a_5}{x^{3/2}\,(\ln{x})^{1-\frac{1}{2}\gamma_m}}\,.
\end{eqnarray}
In principle, the influence of $F_6$, $F_7$, and $F_8$ might change the
power of the logarithm for $F_4$ or $F_5$, but we expect no change in
the leading ultraviolet behavior of $F_1$, $F_2$, and $F_3$.  

\section{Concluding remarks}

We have calculated the light vector meson masses and the decay constants
associated with electromagnetic and leptonic decays
using the ladder truncation for the meson BSE in conjunction with the
rainbow truncation for the quark DSE.  We use an effective
quark-antiquark interaction ${\cal G}(k^2)/k^2$ with one
phenomenological parameter, which is fitted to reproduce $f_\pi$; the
two other parameters are the current quark masses $m_{u/d}$ and $m_{s}$
which are fixed through $m_\pi$ and $m_K$.  The calculated values for
the vector meson masses are within 5\% of the experimental values; the
decay constants are within 10\% of their experimental values.  These
results are fairly robust: they are weakly dependent upon the scale at
which the interaction starts to deviate from the perturbative behavior,
as long as the parameters are fitted to pseudoscalar observables.

An earlier BSE study\cite{JM93} in a related framework produced
qualitatively comparable results for $m_\rho$, $m_{K^\star}$ and
$m_\phi$ as part of a study that included heavy mesons and incorporated
five quark flavors.  Vector meson decay constants were not considered.
That approach produced a dependence upon the momentum-sharing parameter
$\eta$ that is stronger than what we find.  The present results for
physical observables, such as the mass and decay constant, are {\em
independent} of the momentum sharing, as long as all relevant covariants
and the full angular dependence are included in the calculation.  A
recent work~\cite{kissl} has explored the feasibility of extracting
ground state vector meson masses from the large Euclidean time behavior
of the quark current-current correlator as calculated from the ladder
truncation of the inhomogeneous BSE for the vector vertex.  Only the
$\rho$ was studied, the BS amplitudes were not extracted and the decay
constant was not calculated.

Of the eight allowed transverse covariants, five are both quantitatively
and qualitatively important, whereas the remaining three amplitudes
contribute little to the mass and decay constant.  Neglect of these
three amplitudes changes the calculated masses by only 2\% and decay
constants by 8\%.  For the $\rho$ and $\phi$ the dependence of the BS
amplitudes on $q\cdot P$ is very small; truncation to the leading
Chebyshev moments leads to very similar results.  However, the second
moment of $F_1$ is needed for convergence of the loop integral for the
decay constant.  This suggests that, in general, hadronic observables
can be well-described by a rather limited number of covariants and
Chebyshev moments.  For the $K^\star$ however, more Chebyshev moments
are required, since it is not a charge conjugation eigenstate.  Our
numerical results can be used to guide the development of approximating
forms for the BS amplitudes for calculation of a variety of observables
such as electromagnetic form factors and strong decays typified by as
\mbox{$\rho \rightarrow \pi\pi$} and \mbox{$\phi \rightarrow KK$}.

The ladder truncation of the BSE is known to be a good approximation for
flavor nonsinglet pseudoscalar mesons~\cite{BRvS96}, and it is expected
to be reliable for vector mesons as well.  This is to be contrasted with
the scalar channel, where the same analysis revealed~\cite{R97ladsc} that
the next-order corrections are much more important.  For flavor-singlet
mesons, there are also contributions from diagrams corresponding to
quark annihilation to time-like gluons. These play an important role for
pseudoscalars, e.g. in the generation of the $\eta'$ mass through the
axial anomaly~\cite{etamass}.  For the flavor-singlet vector mesons
however, there is no such anomaly.  Also, if quark annihilation diagrams
play a major role for such mesons one would expect more flavor mixing
than is evident for the $\omega$ and $\phi$.  It is therefore reasonable
to expect the ladder truncation to be appropriate for vector mesons.
For the ground state vector mesons considered here, there is an open
decay channel to a pair of pseudoscalars (e.g. \mbox{$\rho\rightarrow
\pi\pi$}), but this is a P-wave coupling that tends to suppress the
mechanism relative to such a decay of a scalar.  Estimates of the
effects of meson loops on the $\rho$ mass vary between
2\% and 10\%~\cite{rhomeslp}.  With the BS amplitudes
calculated here we expect to be able to investigate the effects of meson
dressing more accurately in the future.  Note that both the meson
dressing and the quark annihilation diagrams can contribute to the
splitting between the $\rho$ and $\omega$, which are degenerate in the
ladder truncation.

The task of modeling vector mesons within QCD at finite temperature and 
chemical potential has recently begun with extremely simplified Ans\"atze 
for the kernel of the BSE~\cite{MRS98BT99}.   The present work may provide 
valuable guidance
for the extension and improvement of such efforts to explore the behavior
of vector $\bar q q$ states and correlations relevant to chiral restoration
and quark deconfinement transitions.

\acknowledgements 
We acknowledge useful conversations and correspondence with
C.D.~Roberts, L.S.~Kisslinger, and D.~Jarecke.  This work was funded by
the National Science Foundation under grant No.~PHY97-22429, and
benefited from the resources of the National Energy Research Scientific
Computing Center.

\appendix

\section{Asymptotic behavior of the BS amplitudes}
\label{AppAsRho}

In order to analyze the asymptotic behavior of the BS amplitudes
${}^i\!F_j(q^2;P^2)$, we have to perform all angular integrals
analytically.  These angular integrals have measure
\begin{eqnarray}
\int d\Omega_{p,q} & := & 
        \frac{2}{\pi^2} \int_0^\pi d\theta_p\,\sin^2{\theta_p}\,
        \int_0^\pi d\theta_q\,\sin^2{\theta_q}\,
        \int_0^{2\pi} d\phi\,\sin{\phi} = 1 \,,
\end{eqnarray}
where $\theta_p$ is the angle between the external momentum $p$ and $P$,
and $\theta_q$ is the angle between the integration momentum $q$ and
$P$.  This type of integral can be performed with the help of the
appendix of Ref.~\cite{pmwang}, and some typical results are
\begin{eqnarray}
\int d\Omega_{p,q} \frac{p\cdot q}{(p-q)^2} 
        &=& \frac{p\,q\,\min(p,q)}{2\max(p,q)^3}              \,,\\
\int d\Omega_{p,q} \frac{p\cdot q}{(p-q)^4} 
        &=& \frac{p\,q\,\min(p,q)}
                        {\max(p,q)^3\,(\max(p,q)^2-\min(p,q)^2)}  \,.
\end{eqnarray}  
Other, more complicated, integrals can be expressed in a similar way.  A
common feature of these angular integrals is that they can all be
expressed in terms of $\max(p,q)$ and $\min(p,q)$.  This allows us to
convert the integral equations to differential equations~\cite{atkjohn}.

\begin{figure}
\centering{\
\epsfig{figure=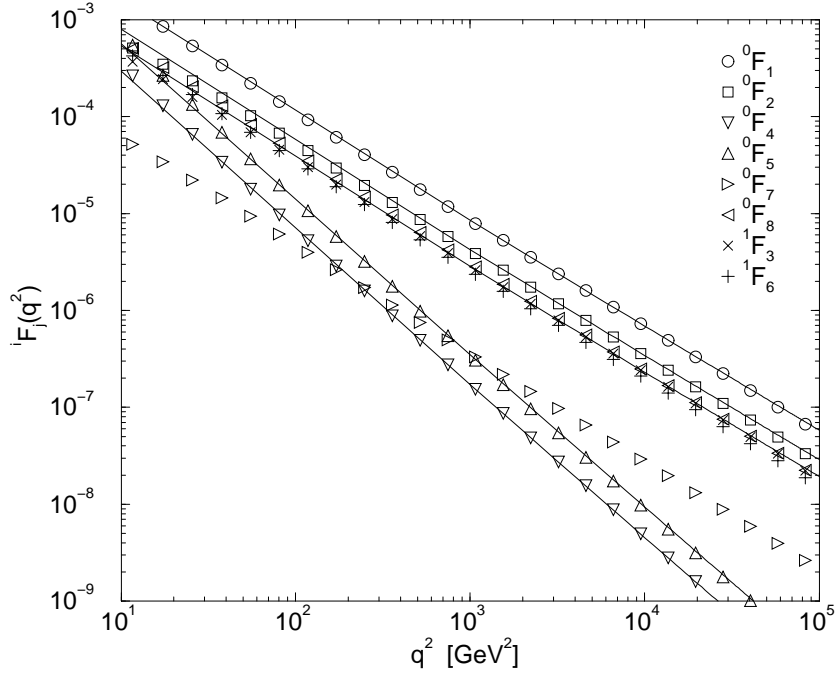,height=9.0cm}}
\caption{
The ultraviolet behavior of the $\rho$ BS amplitudes: the leading
Chebyshev moments of $F_1$-$F_8$, obtained using the full angular
dependence, are shown by the symbols.  The lines display the
analytically calculated behavior for $F_1$-$F_5$, given by
Eqs.~(\ref{UVBSA1})-(\ref{UVBSA3}), (\ref{UVBSA4}), and (\ref{UVBSA5}).
\label{figrhoUV}}
\end{figure}

We have performed all the angular integrals in the five coupled integral
equations for ${}^0\!F_1(q^2;P^2)$, ${}^0\!F_2(q^2;P^2)$,
${}^1\!F_3(q^2;P^2)$, ${}^0\!F_4(q^2;P^2)$, and ${}^0\!F_5(q^2;P^2)$,
ignoring the functions $F_6$, $F_7$, and $F_8$, and truncating the
Chebyshev moments at the leading order.  With the leading ultraviolet
behavior of the functions $F_1$, $F_2$, and $F_3$ considered first, the
relevant kernels are
\begin{equation}
\nonumber
\begin{array}{lcccl}
\nonumber
        {\rm K}_{x>y}^{11}(x,y) = \frac{1}{4\,x}        \,,&&&&
        {\rm K}_{y>x}^{11}(x,y) = \frac{x}{4\,y^2}      \,,\\[2mm]
        {\rm K}_{x>y}^{12}(x,y) = \frac{-\sqrt{5}}{6\,x} \,,&&&&
        {\rm K}_{y>x}^{12}(x,y) = \frac{-\sqrt{5}(4\,y-x)}{18\,y^2} 
                                                        \,,\\[2mm]
        {\rm K}_{x>y}^{13}(x,y) = \frac{-1}{6\,x}       \,,&&&&
        {\rm K}_{y>x}^{13}(x,y) = \frac{-(4\,y-x)}{18\,y^2} \,,\\[3mm]
        {\rm K}_{x>y}^{21}(x,y) = \frac{-\sqrt{5}(4\,y-3\,x)}{54\,x^2}  
                                                        \,,&&&&
        {\rm K}_{y>x}^{21}(x,y) = \frac{-\sqrt{5}\;x}{54\,y^2}          
                                                        \,,\\[2mm]
        {\rm K}_{x>y}^{22}(x,y) = \frac{5(9\,y-8\,x)}{216\,x^2} \,,&&&&
        {\rm K}_{y>x}^{22}(x,y) = \frac{5\;x^2}{216\,y^3} \,,\\[2mm]
        {\rm K}_{x>y}^{23}(x,y) = \frac{-\sqrt{5}(8\,x-3\,y)}{216\,x^2} 
                                                        \,,&&&&
        {\rm K}_{y>x}^{23}(x,y) = \frac{-5\sqrt{5}\;x^2}{216\,y^3}      
                                                        \,,\\[3mm]
        {\rm K}_{x>y}^{31}(x,y) = \frac{-(4\,y-3\,x)}{36\,x^2} \,,&&&&
        {\rm K}_{y>x}^{31}(x,y) = \frac{- x}{36\,y^2}   \,,\\[2mm]
        {\rm K}_{x>y}^{32}(x,y) = \frac{-\sqrt{5}(8\,x-3\,y)}{144\,x^2}  
                                                        \,,&&&&
        {\rm K}_{y>x}^{32}(x,y) = \frac{-5\sqrt{5}\;x^2}{144\,y^3}       
                                                        \,,\\[2mm]
        {\rm K}_{x>y}^{33}(x,y) = \frac{17\,y-8\,x}{144\,x^2} \,,&&&&
        {\rm K}_{y>x}^{33}(x,y) = \frac{(25\,x-16\,y)\,x}{144\,y^3} \,.
\nonumber
\end{array}
\nonumber
\end{equation}
These are to be inserted into the integral equation
Eq.~(\ref{genUVinteq}), which is
\begin{eqnarray}
\label{genUVinteqA}
 F_i(x) &=& \frac{\gamma_m}{\ln{x}} 
        \int_0^x dy\; {\rm K}_{x>y}^{ij}(x,y) F_j(y) 
        + \gamma_m \int_x^\infty dy\; {\rm K}_{y>x}^{ij}(x,y) 
        \frac{F_j(y)}{\ln{y}} 
\end{eqnarray}
where $F_i(x) = {}{}^0\!F_i(q^2;P^2)$ for $i = 1,2$ and $F_3(x) =
{}{}^1\!F_3(q^2;P^2)$.  This set of coupled integral equations can now
be converted into a set of coupled fourth-order differential equations
for $F_{1-3}(x)$ of the type
\begin{eqnarray}
\label{genUVdifeq}
        x^4\,\kappa_{4\,i} \,F^{''''}_i(x) 
        + x^3\,\kappa_{3\,i} \,F^{'''}_i(x) 
        + x^2\,\kappa_{2\,i} \,F^{''}_i(x) 
        + x\,\kappa_{1\,i} \,F^{'}_i(x) 
        + \kappa_{0\,i} \,F(x)_i & = & 0 \,.
\end{eqnarray}
Substitution of the series expansion
\begin{eqnarray}
\label{BSasym}
 F_i(x) &=& \frac{a_i\,(\ln{x})^\alpha }{x^\beta}
                \left( 1 + \sum_{j=1}^\infty c_i^j\,(\ln{x})^{-j} \right)
\end{eqnarray}
into the set of differential equations leads to a set of coupled
equations for the powers $\alpha$ and $\beta$, and the leading
coefficients $a_i$.  It is easy to see that all terms in the
differential equation have the same power of $x$, and collection of all
the leading powers of $\ln{x}$ gives an equation for the power $\beta$.
One of the solutions of this equation is $\beta = 1$, which is obviously
the physical solution, see Fig.~\ref{figrhoUV}.  The next-to-leading
order terms lead to three coupled equations for the four constants
$\alpha$, $a_1$, $a_2$, and $a_3$; the homogeneous BSE allows for an
arbitrary overall scaling and we set $a_1 = 1$.  The solution for the
other constants is then
\begin{eqnarray}
        \alpha &=& -1 + \frac{\gamma_m}{108}    \,,       \\
        a_2 &=& \frac{2\,\sqrt{5}}{9}           \,,      \\
        a_3 &=& \frac{1}{3}                     \,.
\end{eqnarray}
Note that the powers $\alpha$ and $\beta$ are the same for all three
functions.  Differences between the functions only arise from
differences in the coefficients $a_i$ for the leading, and the
subleading coefficients $c_i^j$.

Next we consider ${}^0\!F_4$, which decouples from the other amplitudes
after performing the angular integrals.  The only nonzero kernels in
Eq.~(\ref{genUVinteqA}) are
\begin{equation}
\nonumber
\begin{array}{lcccl}
\nonumber
        {\rm K}_{x>y}^{44}(x,y) = \frac{\sqrt{y}}{x^{3/2}}      \,,&&&&
        {\rm K}_{y>x}^{44}(x,y) = \frac{\sqrt{x}}{y^{3/2}}      \,.
\nonumber
\end{array}
\nonumber
\end{equation}
The resulting asymptotic behavior can be expressed by Eq.~(\ref{BSasym}) with
\begin{eqnarray}
        \beta_4  &=& \case{3}{2}                \,\\
        \alpha_4 &=& -1 +\case{1}{3}\gamma_m    \,.
\end{eqnarray}
This is in agreement with the numerical result, see Fig.~\ref{figrhoUV}.
The equation for $F_5$ is more complicated, since ${}^0\!F_5$ does
couple to ${}^0\!F_1$, ${}^0\!F_2$, and ${}^1\!F_3$.  The relevant
kernels are
\begin{equation}
\nonumber
\begin{array}{lcccl}
\nonumber
  {\rm K}_{x>y}^{51}(x,y) = \frac{M(y)}{2\,x^{3/2}}             \,,&&&&
  {\rm K}_{y>x}^{51}(x,y) = \frac{\sqrt{x}M(y)}{2\,y^2}         \,,\\[2mm]
  {\rm K}_{x>y}^{52}(x,y) = \frac{\sqrt{5}M(y)}{3\,x^{3/2}}     \,,&&&&
  {\rm K}_{y>x}^{52}(x,y) = \frac{\sqrt{5}\sqrt{x}M(y)}{3\,y^2} \,,\\[2mm]
  {\rm K}_{x>y}^{53}(x,y) = \frac{M(y)}{3\,x^{3/2}}             \,,&&&&
  {\rm K}_{y>x}^{53}(x,y) = \frac{\sqrt{x}M(y)}{3\,y^2}         \,,\\[2mm]
  {\rm K}_{x>y}^{55}(x,y) = \frac{\sqrt{y}}{2\,x^{3/2}}         \,,&&&&
  {\rm K}_{y>x}^{55}(x,y) = \frac{\sqrt{x}}{2\,y^{3/2}}         \,.
\nonumber
\end{array}
\nonumber
\end{equation}
However, a careful analysis shows that the leading ultraviolet behavior
of ${}^0\!F_5$ is not influenced by coupling to other amplitudes; the
leading behavior arises from $K^{55}$ only.  The result is
Eq.~(\ref{BSasym}) with
\begin{eqnarray}
        \beta_5  &=& \case{3}{2}                \,,\\
        \alpha_5 &=& -1 +\case{1}{2}\gamma_m    \,,
\end{eqnarray}
also in agreement with our numerical results, see Fig.~\ref{figrhoUV}.
The influence of ${}^0\!F_5$ on our previous results for ${}^0\!F_1$,
${}^0\!F_2$, and ${}^1\!F_3$ can be examined for consistency.  Those
three amplitudes fall off like $1/x$, while contributions from
${}^0\!F_5$ to these amplitudes via the differential equations in
Eq.~(\ref{genUVdifeq}) will be suppressed by a factor of
$M(x)/\sqrt{x}$, and thus contribute to the subleading behavior only.

%
%
%
%

%
%
%
%
\newpage
%
%
%
\begin{table}
\begin{tabular}{l|ccccc}
& experiment  &         &$\omega=0.3\,{\rm GeV}$
                                &$\omega=0.4\,{\rm GeV}$
                                        &$\omega=0.5\,{\rm GeV}$\\
& (estimates) & Ref.~\cite{MR97}& $D=1.25\,{\rm GeV}^2$   
                                & $D=0.93\,{\rm GeV}^2$   
                                        & $D=0.79\,{\rm GeV}^2$ \\ \hline
- $\langle \bar q q \rangle^0_{\mu=1 {\rm GeV}}$
                & $(0.236 \rm GeV)^3$ & $(0.241 \rm GeV)^3$ 
                                        & 0.242 & 0.241 & 0.243  \\
$m^{u=d}_{\mu=1 {\rm GeV}}$       
                & 5 - 10 MeV  & 5.5 MeV & 5.54  &  5.54 &  5.35  \\
$m^{s}_{\mu=1 {\rm GeV}}$         
                &100 - 300 MeV& 130 MeV & 124   &  125  &  123   \\
$m_\pi$         &  0.1385 GeV & 0.1385  & 0.139 & 0.138 & 0.138  \\
$f_\pi$         &  0.1307 GeV & 0.1307  & 0.130 & 0.131 & 0.131  \\
$m_K$           &  0.496 GeV  & 0.497   & 0.496 & 0.497 & 0.497  \\
$f_K$           &  0.160 GeV  & 0.154   & 0.154 & 0.155 & 0.157 
\end{tabular}
\caption{\label{respseudo}
Calculated values of the properties of light, pseudoscalar mesons, for
the parameterization of the effective interaction Eq.~(\protect\ref{gvk2}),
using three different parameter sets, and also for the parameterization
of Ref.~\protect\cite{MR97}.}
\end{table}
\begin{table}
\begin{tabular}{l|dddddd}
 & \multicolumn{2}{c}{$\rho$}      & 
   \multicolumn{2}{c}{$K^\star$}   &
   \multicolumn{2}{c}{$\phi$}      \\ 
                & $m_\rho$ & $f_\rho$ 
                & $m_{K^\star}$ & $f_{K^\star}$ & $m_\phi$ & $f_\phi$ \\ \hline
experiment      & 0.770 & 0.216 & 0.892 & 0.225 & 1.020 & 0.236 \\ \hline
All amplitudes $F_1$-$F_8$  
                &       &       &       &       &       &       \\ \hline
$\omega=0.3\,{\rm GeV}$, $D=1.20\,{\rm GeV}^2$  
                & 0.747 & 0.197 & 0.956 & 0.246 & 1.088 & 0.255 \\
$\omega=0.4\,{\rm GeV}$, $D=0.93\,{\rm GeV}^2$ 
                & 0.742 & 0.207 & 0.936 & 0.241 & 1.072 & 0.259 \\
$\omega=0.5\,{\rm GeV}$, $D=0.79\,{\rm GeV}^2$ 
                & 0.74  & 0.215 & 0.94  & 0.25  & 1.08  & 0.266 \\ \hline
\multicolumn{7}{l}{amplitudes $F_1 \ldots F_5$ only}            \\ \hline
Maris--Roberts Ref.~\cite{MR97} 
                & 0.71&       & 0.95  &       & 1.1   & \\ \hline
$\omega=0.3\,{\rm GeV}$, $D=1.20\,{\rm GeV}^2$  
                & 0.737 & 0.192 & 0.942 & 0.235 & 1.080 & 0.247 \\
$\omega=0.4\,{\rm GeV}$, $D=0.93\,{\rm GeV}^2$ 
                & 0.729 & 0.199 & 0.919 & 0.229 & 1.062 & 0.250 \\
$\omega=0.5\,{\rm GeV}$, $D=0.79\,{\rm GeV}^2$ 
                & 0.731 & 0.207 & 0.926 & 0.237 & 1.072 & 0.259 
\end{tabular}
\caption{\label{resvecvarious}
Comparison of the results for the vector mesons for the three different
parameter sets for the effective interaction, using all eight BS
amplitudes (top), and using the five leading BS amplitudes only
(bottom).}
\end{table}
\newpage
\begin{table}[b]
\begin{tabular}{l|dddddd}
full angular &\multicolumn{2}{c}{$\rho$} & \multicolumn{2}{c}{$K^\star$} &
                                        \multicolumn{2}{c}{$\phi$}      \\ 
calculation  & $m_\rho$ & $f_\rho$ & $m_{K^\star}$ 
                        & $f_{K^\star}$ & $m_\phi$ & $f_\phi$   \\ \hline
all 8 amplitudes        & 0.742 & 0.207 & 0.936 & 0.241 & 1.072 & 0.259 \\
$F_1$ only              & 0.88  & 0.20  & $>$ 1.2 & --- & 1.24  & 0.20  \\
$F_1$, $F_2$, and $F_3$ & 0.90  & 0.17  & $>$ 1.2 & --- & 1.25  & 0.20  \\
$F_1$, $F_4$, and $F_5$ & 0.722 & 0.23  & 0.911 & 0.26  & 1.059 & 0.28  \\
$F_1$ \ldots $F_5$      & 0.729 & 0.199 & 0.919 & 0.23  & 1.062 & 0.250 \\ \hline
\multicolumn{7}{l}{leading Chebyshev decomposition}        \\ \hline
all 8 amplitudes        & 0.743 & 0.211 & 0.92  & 0.24  & 1.074 & 0.262 \\
$F_1$ only              & 0.875 & 0.20  & 1.09  & 0.22  & 1.24  & 0.22  \\
$F_1$, $F_2$, and $F_3$ & 0.900 & 0.17  & 1.10  &  --   & 1.25  & 0.20  \\
$F_1$, $F_4$, and $F_5$ & 0.724 & 0.23  & 0.90  & 0.26  & 1.062 & 0.28  \\
$F_1$ \ldots $F_5$      & 0.730 & 0.201 & 0.91  & 0.23  & 1.065 & 0.251 
\end{tabular}
\caption{\label{resvecnew}
The influence of the different covariants and of the angular dependence
of the amplitudes on the vector meson properties with parameter set
$\omega=0.4\,{\rm GeV}$, $D = 0.93\,{\rm GeV}^2$.  For this table, we
have calculated the loop for the decay constant up to the
renormalization point $\mu=19$~GeV, since for some of the approximations
considered this integral is ultraviolet divergent.  In the case of a
convergent integral, the error made by cutting off the integral at the
renormalization point is less than 1\%.}
\end{table}
%
%
\end{document}